\documentclass[fleqn,twoside]{article}
\usepackage{espcrc2}

\usepackage{graphicx}
\usepackage[figuresright]{rotating}

\newcommand{\AmS}{{\protect\the\textfont2
  A\kern-.1667em\lower.5ex\hbox{M}\kern-.125emS}}

\title{Non-compact QED$_3$ coupled to a four-fermi interaction}

\author{J.B. Kogut \address[MCSD]{Physics Department, University of Illinois, 1110 West Green Street, Urbana, 
IL 61801-3080, USA.},
C.G. Strouthos  \address{Division of Science and Engineering, 
        Frederick Institute of Technology, 
        Nicosia 1303, Cyprus.} and
I.N. Tziligakis \addressmark[MCSD]
}

\begin{document}

\begin{abstract}

We present preliminary numerical results for the three dimensional non-compact
QED (QED$_3$) with a weak four-fermion term in the lattice action. Approaches based on
Schwinger-Dyson studies, arguments based on thermodynamic inequalities and numerical simulations 
lead to estimates of the critical number of fermion flavors (below which chiral symmetry is broken)
ranging from $N_{fc}=1$ to $N_{fc}=4$. 
The weak four-fermion coupling provides the framework for
an improved algorithm, which allows us to simulate the chiral limit
of massless fermions and expose delicate effects.

\vspace{1pc}
\end{abstract}

\maketitle

Three dimensional QED is an interesting and challenging field theory
which serves as a laboratory to study dynamical mass generation.
It is confining and it is believed to exhibit chiral symmetry breaking when the number of fermion 
flavors $N_f$ is smaller than a critical value $N_{fc}$. 
The model is super-renormalizable, because the gauge coupling has dimension $\frac{1}{2}$ 
and so provides an intrinsic scale of the theory. 
Interest in this theory has recently been revived by suggestions that QED$_3$ may be an effective theory 
for the  underdoped and non-superconducting region of the phase diagram of high-$T_c$ superconducting 
cuprate compounds \cite{supercond}. Preliminary numerical results of QED$_3$ with Fermi and gap anisotropies were
presented at this conference \cite{iori}.

The principal analytical approach to study dynamical symmetry breaking in QED$_3$ is the self-consistent solution 
of truncated Schwinger-Dyson equations for the fermion propagator. Most recent attempts  yielded $N_{fc} \simeq 4$ 
\cite{sde}. An argument based on a thermodynamic inequality yields the prediction $N_{fc} \leq \frac{3}{2}$ \cite{appelquist}. 
Recent attempts to decide the issue by numerical simulations placed an upper bound on the dimensionless
condensate $\beta^2 \langle \bar{\psi} \psi \rangle$ of $O(10^{-4})$ for $N_f=2$ \cite{kogut2002}. 
It was also shown that 
for $N_f=1$ the $\beta^2 \langle \bar{\psi} \psi \rangle$ is of $O(10^{-3})$ but it appeared difficult
to determine whether $N_f=4$ lies above or below $N_{fc}$ \cite{kogut2004}. 

Our purpose is to develop firmer control over the computer simulation studies of chiral 
symmetry breaking by studying the theory with massless fermions. 
Dealing with almost massless fermions, as must be done in the search
of continuum and chiral limits of lattice models is very difficult with the standard algorithms.
However, it has been shown that when one introduces in $(3+1)d$ QED a four-fermion interaction in a semi-bosonized
fashion the performance of
the algorithm improves dramatically \cite{qed4}. The auxiliary field $\sigma(x)$ acts as a site dependent dynamical
mass term in the fermion propagator. In the broken phase, the Dirac operator is now non-singular for
fermions with zero bare mass and its inversion is accurate and very fast.

We have chosen to add to the QED$_3$ lagrangian a discrete $Z_2$ chiral invariant four-fermion term, which is
preferable over four-fermion terms with continuous chiral symmetry because the latter are not 
as efficiently simulated due to massless modes in the strongly cut-off theory.
The lattice action using staggered lattice fermion fields
$\chi,\bar{\chi}$
is given by the formulas below:
\small{
\begin{eqnarray*}
S&=&\frac{\beta}{2}\! \sum_{x,\mu<\nu}\! F_{\mu \nu}(x)\! F^{\mu \nu}(x)\!  \nonumber \\
&+& \sum_{x,x^\prime} {\bar \chi}(x) Q(x,x^\prime) \chi(x^\prime) 
+\frac{N_f \beta_s}{4} \sum_{\tilde{x}} \sigma^2(\tilde{x}), \nonumber \\
{\rm where} \\
F_{\mu \nu}(x) &\equiv& \alpha_\mu(x)+\alpha_\nu(x+\hat \mu)
-\alpha_\mu(x+\hat \nu)-\alpha_\nu(x), \nonumber  \\
Q(x,x^\prime)&\equiv& 
\frac{1}{2}\! \sum_\mu\eta_{\mu}(x)
[\delta_{x^\prime,x+\hat \mu} U_{x\mu}
\!-\!\delta_{x^\prime,x-\hat \mu} U_{x-\hat \mu,\mu}^\dagger] \nonumber  \\
&+& \delta_{xx^{\prime}}\frac{1}{8}\sum_{\langle \tilde{x}, x \rangle} \sigma(\tilde{x}).
\end{eqnarray*}
}
The indices $x,~x^\prime$ consist of three integers
$(x_1,~x_2,~x_3)$
labelling the lattice sites, where the third direction is now considered
to be timelike.
The symbol $\langle \tilde{x}, x \rangle$ denotes the set of the 8 dual lattice sites $\tilde{x}$
surrounding the direct lattice site $x$.
Since the gauge action $F^2$ is unbounded from above,
this action
defines the {\sl non-compact\/} formulation of lattice QED.
The $\eta_\mu(x)$ are the Kawamoto-Smit staggered fermion phases $(-1)^{x_1+\cdots+x_{\mu-1}}$,
designed to ensure relativistic covariance of the Dirac equation in the continuum limit.
Antiperiodic boundary conditions are used in
the timelike direction and periodic
boundary conditions in the spatial directions for the fermion fields.
The phase factors in the fermion bilinear are defined by
$U_{x\mu} \equiv
\exp(i\alpha_{x \mu})$,
where
$\alpha_{x \mu}$ is the gauge
potential.
In terms of continuum quantities,
$\alpha_{x\mu}= a gA_\mu(x)$, 
$\beta \equiv \frac{1}{g^2 a}$, $\beta_s \equiv \frac{a}{g_s^2}$
where $a$ is the physical lattice spacing, $g_s^2$ is the four-fermion coupling 
and $g^2$ is the gauge coupling.
The numerical results presented here were obtained by simulating the
QED$_3$ action using a standard Hybrid Molecular Dynamics algorithm.
\begin{figure}[t!]
\centering
\includegraphics[scale=0.59]{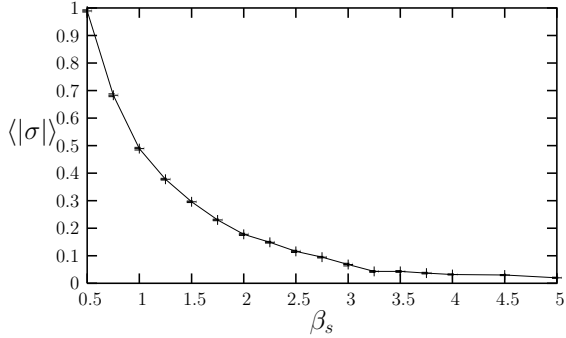}
\caption{Order parameter for $N_f=4$ versus $\beta_s$ at fixed $\beta=0.2$. The lattice size is $12^3$.}
\label{fig:fig1}
\end{figure}
\begin{figure}[t!]
\centering
\includegraphics[scale=0.59]{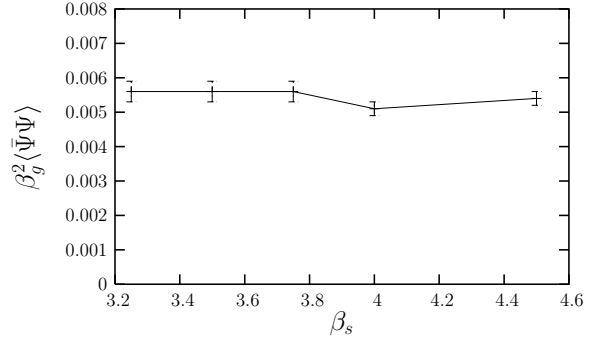}
\caption{Dimensionless condensate for $N_f=4$ versus $\beta_s$ at fixed $\beta=0.2$.}
\label{fig:fig2}
\end{figure}

The vacuum expectation value of the auxiliary field $\sigma$ is the chiral order parameter.
In the strong gauge coupling limit $\beta \rightarrow 0$, it is known rigorously that chiral symmetry 
is broken \cite{salmhofer}. Therefore, as $\beta$ increases, for $N_f > N_{fc}$ there must be a chiral symmetry 
restoring phase transition for some finite $\beta_{c}$. 
A mean field form fitting the data would indicate that our data are compatible with the hypothesis
that the continuum limit of the theory at the critical point describes a free field. Finding evidence
for an interacting continuum model would be very puzzling indeed.
For $N_f< N_{fc}$, since we believe the order parameter
is exponentially small in the continuum limit, the relic of the transition may persist as a crossover
between weak and strong coupling behavior with a tail in $\bar\psi\psi$ extending to weak gauge coupling. 

\begin{figure}[t]
\centering
\includegraphics[scale=0.59]{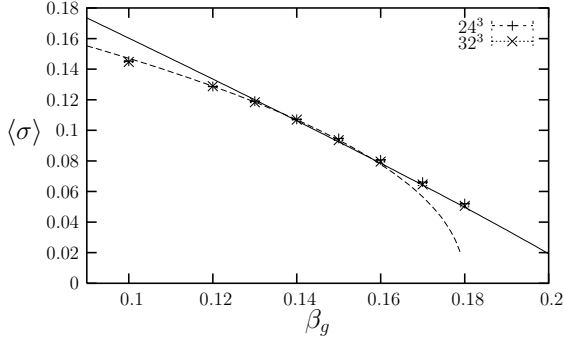}
\caption{Order parameter for $N_f=4$ versus $\beta$ at fixed $\beta_s=4.0$.} 
\label{fig:fig3}
\end{figure}

First, we performed simulations with $N_f=4$. In Fig.~\ref{fig:fig1} we plot $\langle |\sigma| \rangle$ 
versus $\beta_s$ at fixed $\beta=0.2$. The simulations were performed on $12^3$ lattices. 
We should note that in the zero gauge coupling limit the $N_f=4$ three-dimensional four-fermion model undergoes 
a phase transition at $\beta_s=0.86(1)$ \cite{bkt}. 
In the weak four-fermi 
coupling limit all the dynamics are expected to reside in the gauge and fermion fields and the extra four-fermion term 
just provides the framework for an improved algorithm. 
This becomes clearer in Fig.~\ref{fig:fig2} where we see
that the dimensionless condensate $\beta^2 \langle \bar \psi \psi \rangle$ at $\beta=0.2$ in the weak four-fermion 
coupling limit saturates to a nonzero 
value $\approx 0.0055$. This result is consistent with results reported in \cite{kogut2004} which were obtained using standard
chiral extrapolations. In Fig.~\ref{fig:fig3} we show data generated on $24^3$ and $32^3$ lattices 
for $\langle \sigma \rangle$ versus $\beta$ at fixed
$\beta_s=4.0$. The difference between the $24^3$ and the $32^3$ data is small, implying that the $32^3$ data are
close to the thermodynamic limit. 
The $32^3$ data can be fit to the standard scaling relation
$\langle \sigma \rangle = {\rm const}.(\beta_{c}-\beta)^{\beta_{\rm mag}}$.
For $0.12 \leq \beta \leq 0.16$ we get $\beta_{\rm mag}=0.46(3)$ which is consistent with mean field 
theory. For $0.15 \leq \beta \leq 0.18$ we get $\beta_{\rm mag}=0.92(11)$ and $\beta_{c}=0.211(6)$. 
If this value of $\beta_{\rm mag}$ persists for couplings arbitrarily close to $\beta_{c}$ it will become 
clear that the theory has an interacting fixed point. 
A detailed finite size scaling analysis is required to clarify this issue.
However, we should note that this value of $\beta_{\rm mag}$ 
is close to its value ($\beta_{\rm mag}=1.1(1)$ at the UV fixed point of the $(2+1)d$ four-fermion model \cite{bkt}.
Although $\beta_s=4.0$ is deep in the symmetric phase of the ``pure'' four-fermion model, we still have to study
in detail the interplay of the QED$_3$ and the four-fermion fixed points before we reach a conclusion. 
The fact that QED$_3$ is super-renormalizable
may imply that the four-fermion fixed point has a much larger domain of attraction than the QED$_3$ fixed point.
Therefore, we may need to perform simulations with smaller $\beta_s$ before we are certain that
the simulations lie in the vicinity of attraction of the QED$_3$ fixed point.

\begin{figure}[t]
\centering
\includegraphics[scale=0.59]{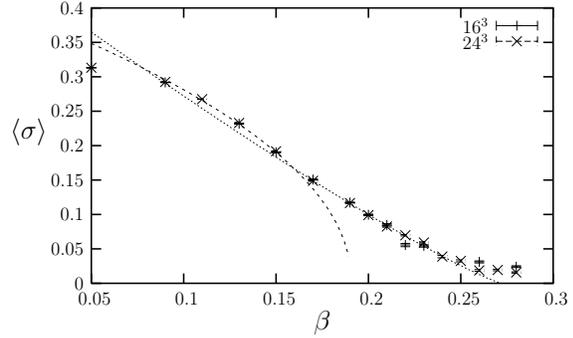}
\caption{Order parameter for $N_f=8$ versus $\beta$ at fixed $\beta_s=2.0$.}
\label{fig:fig4}
\end{figure}
We also performed simulations with $N_f=8$. With fixed $\beta_s=2.0$ we have the following results 
from $24^3$ lattices: 
For $\beta=0.11,0.13,0.15$ we get $\beta_{\rm mag}=0.49(4)$ which is consistent with mean field theory. 
However, for $\beta=0.17,...,0.24$ we get $\beta_{\rm mag}=1.15(8)$ and $\beta_c=0.272(5)$. This value of
$\beta_{\rm mag}$ is close to its value in the $(2+1)d$ four-fermion model. We are currently 
performing simulations with $\beta_s=4.0$. Although our results are very preliminary, the data provide 
some evidence that mean field theory scaling is valid in a region where the values of $\langle \sigma \rangle$
are smaller than in the mean field theory region of the $\beta_s=2.0$ case. In the mean field region $\langle \sigma \rangle$
is equal to the fermion mass. It will be interesting to check whether this scenario persists
at values of $\beta$ arbitrarilly close to the transition or whether a crossover to a non-trivial 
scaling region occurs at smaller values of $\langle \sigma \rangle$.

\end{document}